\documentstyle[12pt,graphicx,subfigure,epsfig,epsf]{article}
\def\slash#1{\setbox0=\hbox{$#1$}#1\hskip-\wd0\hbox to\wd0{\hss\sl/\/\hss}}
\begin{document}
\baselineskip=20 pt
\def\l{\lambda}
\def\m{\mu}
\def\L{\Lambda}
\def\bt{\beta}
\def\mphi{m_{\phi}}
\def\hphi{\hat{\phi}}
\def\vphi{\langle \phi \rangle}
\def\etamunu{\eta^{\mu\nu}}
\def\dmul{\partial_{\mu}}
\def\dnul{\partial_{\nu}}
\def\bea{\begin{eqnarray}}
\def\eea{\end{eqnarray}}
\def\bfl{\begin{flushleft}}
\def\efl{\end{flushleft}}
\def\bc{\begin{center}}
\def\ec{\end{center}}
\def\bcr{\begin{center}}
\def\ecr{\end{center}}
\def\al{\alpha}
\def\bt{\beta}
\def\eps{\epsilon}
\def\lam{\lambda}
\def\gam{\gamma}
\def\s{\sigma}
\def\r{\rho}
\def\e{\eta}
\def\dl{\delta}
\def\non{\nonumber}
\def\nont{\noindent}
\def\la{\langle}
\def\ra{\rangle}
\def\nc{{N_c^{\rm eff}}}
\def\vp{\varepsilon}
\def\drho{\bar\rho}
\def\deta{\bar\eta}
\def\vma{{_{V-A}}}
\def\vpa{{_{V+A}}}
\def\J{{J/\psi}}
\def\ov{\overline}
\def\Lqcd{{\Lambda_{\rm QCD}}}
\def\btr{\bigtraingleup}


\def\issue(#1,#2,#3){#1 (#3) #2} 
\def\APP(#1,#2,#3){Acta Phys.\ Polon.\ \issue(#1,#2,#3)}
\def\ARNPS(#1,#2,#3){Ann.\ Rev.\ Nucl.\ Part.\ Sci.\ \issue(#1,#2,#3)}
\def\CPC(#1,#2,#3){Comp.\ Phys.\ Comm.\ \issue(#1,#2,#3)}
\def\CIP(#1,#2,#3){Comput.\ Phys.\ \issue(#1,#2,#3)}
\def\EPJC(#1,#2,#3){Eur.\ Phys.\ J.\ C\ \issue(#1,#2,#3)}
\def\EPJD(#1,#2,#3){Eur.\ Phys.\ J. Direct\ C\ \issue(#1,#2,#3)}
\def\IEEETNS(#1,#2,#3){IEEE Trans.\ Nucl.\ Sci.\ \issue(#1,#2,#3)}
\def\IJMP(#1,#2,#3){Int.\ J.\ Mod.\ Phys. \issue(#1,#2,#3)}
\def\JHEP(#1,#2,#3){J.\ High Energy Physics \issue(#1,#2,#3)}
\def\JPG(#1,#2,#3){J.\ Phys.\ G \issue(#1,#2,#3)}
\def\MPL(#1,#2,#3){Mod.\ Phys.\ Lett.\ \issue(#1,#2,#3)}
\def\NP(#1,#2,#3){Nucl.\ Phys.\ \issue(#1,#2,#3)}
\def\NIM(#1,#2,#3){Nucl.\ Instrum.\ Meth.\ \issue(#1,#2,#3)}
\def\PL(#1,#2,#3){Phys.\ Lett.\ \issue(#1,#2,#3)}
\def\PRD(#1,#2,#3){Phys.\ Rev.\ D \issue(#1,#2,#3)}
\def\PRL(#1,#2,#3){Phys.\ Rev.\ Lett.\ \issue(#1,#2,#3)}
\def\PTP(#1,#2,#3){Progs.\ Theo.\ Phys. \ \issue(#1,#2,#3)}
\def\RMP(#1,#2,#3){Rev.\ Mod.\ Phys.\ \issue(#1,#2,#3)}
\def\SJNP(#1,#2,#3){Sov.\ J. Nucl.\ Phys.\ \issue(#1,#2,#3)}
\def\ZPC(#1,#2,#3){Zeit.\ Phys.\ C \issue(#1,#2,#3)}

\def\pr{{\sl Phys. Rev.}~}
\def\prl{{\sl Phys. Rev. Lett.}~}
\def\pl{{\sl Phys. Lett.}~}
\def\np{{\sl Nucl. Phys.}~}
\def\zp{{\sl Z. Phys.}~}

\font\el=cmbx10 scaled \magstep2{\obeylines\hfill IMSc-2006/05/22}

\begin{center}

{\large \bf ${\overline B_s} \to \mu^+ \mu^-$ decay in the Randall-Sundrum model} \\
\end{center}

\vskip 10pt

\begin{center}
{\sl \large {Basudha Misra~\footnote{E-mail: basudha@imsc.res.in}, Jyoti Prasad Saha~\footnote{E-mail: jyoti@imsc.res.in} \\  and \\ Prasanta Kumar Das~\footnote{E-mail: dasp@imsc.res.in}}}
\end{center}
\vskip  5pt
\begin{center}
The Institute of Mathematical Sciences,\\
C.I.T Campus, Taramani,\\
Chennai 600113, India.
\end{center}

\centerline{\bf Abstract} {\small We investigate the 
${\overline B_s} \to \mu^+ \mu^-$ 
decay in the presence of a light stabilized radion in Randall-Sundrum model. 
The branching ratio 
$BR({\overline B_s} \to \mu^+ \mu^-)$ in 
the standard model is found to be $3.17 \times 10^{-9}$ 
(two order smaller than the 
experimental upper bound) and raises the question whether some new physics 
can play a crucial role or not. We found that for a 
reasonable range of parameters (i.e. radion mass $m_\phi$ and radion vev 
$\vphi$), the above deficit between the experimental bound and the standard 
model result can well be accomodated. Using this upper
bound on $BR({\overline B_s} \to \mu^+ \mu^-)$, we obtain 
the lower bound on $\vphi$.   

\bfl
\vspace*{0.05in}
\efl

\newpage
\section{Introduction}
The standard model(SM) despite it's enormous experimental success, succumbs 
some serious drawbacks as far as our understandings of the flavour structure of 
the quarks and leptons are concerned. The CKM matrix, telling a lot about 
the mixing and CP-violation in the quark sector, lacks any dynamical mechanism in it's origin. Simlilarly, the neutrino masses and mixing which is absent in 
the SM, requires 
the existence of beyond SM physics. An urge for going beyond the SM (by invoking new physics(NP)) has become a driving force behind most of the present phenomenological studies.  Ideas like supersymmetry (with or without 
$R$-parity), technicolor, extra-dimensional field theory, split fermion as a 
candidate of NP draw a lot of attention among the physics community and 
the result is that the LANL archive is flourished by thousands of paper in each year. Among several ingenious ideas the notion of extra spatial dimension(s)
(\cite{ADD}, \cite{RS}) have become very popular as they partially resolve hierarchy puzzle. Among these the 
Randall-Sundrum(RS) model (of warped extra spatial dimension ) 
is particularly interesting from the phenomenological point of view. This model views the world 
as $5$-dimensional and it's fifth spatial dimension is 
$S^1/Z_2$ orbifold. The metric describing such a world can be written as
\bea
d s^2 = \Omega^2 \eta_{\mu \nu} d x^\mu d x^\nu
- R_c^2 d \theta^2.
\eea
\noindent
The pre-factor $\Omega^2 = e^{-2 k R_c |\theta|}$ appearing in above  
is called the warp factor. In $\Omega^2$, $k$ stands for the bulk curvature constant and $R_c$ for the size of the extra dimension. The angular variable 
$\theta$ parametrizes the fifth dimension ($x_5 = R_c \theta$). 
The model is constructed out of  
two $D_3$ branes located at the two orbifold fixed points 
$\theta = 0$ and $\theta = \pi$, respectively. The brane located at  
$\theta = 0$ (where gravity peaks) is known as the Planck brane, whereas  
the brane located at  $\theta = \pi$ (on which the SM fields resides and the gravity becomes weak) is called the TeV brane. 

The radius $R_c$ can be 
related to the vacuum expectation value (vev) of some modulus
field $T(x)$ which corresponds to the fluctuations of the metric over the
background geometry (given by $R_c$). Replacing $R_c$ by $T(x)$ we can rewrite
the RS metric at the orbifold point $\theta = \pi$ as
\bea
d s^2 = g_{\mu \nu}^{vis} d x^\mu d x^\nu, 
\eea

\noindent
where $g_{\mu \nu}^{vis} = e^{- 2 \pi k T(x)}\eta_{\mu \nu}
= \left(\frac{\phi(x)}{f}\right)^2 \eta_{\mu \nu}$. Here
$f^2 = \frac{24 M_5^3}{K}$, where $M_5$ is the $5$-dimensional Planck scale
\cite{RS} and $\phi(x) = f e^{- \pi k T(x)}$. The scalar field $\hat{\phi}(x)$ 
(i.e. $\hat{\phi}(x) = \phi(x) - \vphi $) is known as the radion 
field. In the minimal version of the RS model there is no potential which can stabilize the modulus field $T(x)$ (and thus the radion $\hat{\phi}(x)$). However in a pioneering work, Goldberger and Wise \cite{GW} were able to 
generate a potential of this modulus field (by adding an extra massive bulk 
scalar) field which has the correct minima satisfying $k R_{c} \simeq 11-12$, a necessary condition for the hierarchy resolution. 

In this non-minimal RS model (RS model together with the Goldberger and Wise mechanism), the stabilzed radion can be lighter than the 
other low-lying gravitonic degrees of freedom and will reveal  
itself first either in the direct collider search 
or indirectly through the precission measurement. Studies based on the 
observable consequences of radion are available in the literature 
(\cite{DM}).

 However, the impact of radion in heavy $B$ meson decay, particularly in the rare decay mode, is not fully explored and the present work is an effort in that direction. We have investigated the  rare ${\overline B_s} \to \mu^+ \mu^-$ mode in the light of a stabilized radion. 
 The branching ratio  $BR({\overline B_s} \to \mu^+ \mu^-)$ within the SM is found to be two order smaller than the experimental upper bound $1.0 \times 10^{-7}$ 
 \cite{HFAG}. It is worthwhile to see whether this deficit can be accommodated within this brane world model.

The organization of the paper is as follows. 
In Section~\ref{sec:section2} we obtain the $BR({\overline B_s} \to \mu^+ \mu^-)$ within the standard model and in the presence of a light stabilized radion.  
We discuss several input parameters: CKM matrix elements, quark and lepton masses, effective wilson coefficients, decay constants in 
Section~\ref{sec:section3}.  Section~\ref{sec:section4} is fully devoted to  
the numerical analysis. Here we show how the upper bound of $BR({\overline B_s} \to \mu^+ \mu^-)$ can give some lower bound on radion vev $\vphi$ for a relatively light radion. We summarize and conclude in Section~\ref{sec:section5}.

\section{The exclusive rare ${\overline B_s} \to \mu^+ \mu^-$ decay}
\label{sec:section2}

\subsection{Standard Model(SM)}
The ${\overline B_s}(p_B) \to \mu^+(p_1) \mu^-(p_2)$ (partonically described 
as the $b \overline{s} \to \mu^+ \mu^-$ FCNC transition) decay within the SM is found to be two order smaller than the experimental upper bound. 
This channel can be useful to test certain class of New Physics, particularly 
the notion of warped geometry by probing it's moduli (radion). 

To begin with let us recall the result of the 
${\overline B_s} \to \mu^+ \mu^-$ decay in the standard model \cite{Buras}.
The QCD improved effective Hamiltonian $ {\cal H}_{\rm eff}$ describing such a 
$\Delta B = 1$ transition can be written as
 \bea\label{eqn:eff}
 {\cal H}_{\rm eff} = \frac{G_F}{\sqrt{2}} \left[ V_{tb} V_{ts}^* \left(  c^{eff}_7 O_7 +c^{eff}_{9} O_{9} + c_{10} O_{10} \right)\right], 
\eea
where the operators $O_{i}(~i=9,10)$ (semileptonic operators involving 
electro-weak ($\gamma, Z$) penguin and box diagram) 
and $O_7$ (magnetic penguin) \cite{Buras,Ali} are given by,
\bea \label{eqn:effopr}
 O_9 &=& \frac{\alpha}{\pi} ({\overline{s}} \gamma_\mu P_L b) ({\overline{\mu}}^- \gamma^\mu \mu^+),
 \nonumber \\
 O_{10} &=& \frac{\alpha}{\pi} ({\overline{s}} \gamma_\mu P_L b) ({\overline{\mu}}^- \gamma^\mu \gamma_5 \mu^+),
  \nonumber \\
  O_{7} &=& \frac{\alpha}{\pi} ({\overline{s}} \sigma_{\mu\nu} q^\nu P_R b) \left[\frac{- 2 i m_b}{q^2}\right]({\overline{\mu}}^- \gamma^\mu \mu^+).
  \eea

Here $\sigma_{\mu\nu} = \frac{i}{2} \left[\gamma_\mu,\gamma_\nu\right]$, the 
chiral-projection operators $P_{R,L} = \frac{1}{2} (1 \pm \gamma_5)$. 
$\alpha = \frac{e^2}{4 \pi}$ is the QED fine structure constant. 
In above $q$ is the momentum transferred to the lepton pair,
$m_b$, the $b$-quark mass. Since we are considering the 
${\overline B_s}$ meson, the matrix element of ${\cal H}_{\rm eff}$ is to be taken between the vacuum and $|{\overline B_s}\rangle$ state. Defining 
$f_{{\overline B_s}}$ as the decay constant of the ${\overline B_s}$ meson, we find \cite{Ali}
\bea \label{eqn:mesondecay1}
\langle 0 |{\overline s} \gam^\mu \gam^5 b|{\overline B_s} (p_B)\rangle
= i f_{{\overline B_s}} p_{B}^\mu, \\
\langle 0 |{\overline s} \gam^5 b|{\overline B_s} (p_B)\rangle
= \frac{-i f_{\overline B_s} m_{\overline B_s}^2}{m_b + m_s},  
\eea
and
\bea \label{eqn:mesondecay2}
\langle 0 |{\overline s} \sigma^{\mu \nu} P_R b|{\overline B_s} (p_B)\rangle = 0 .
\eea
Since $q=p_1 + p_2$, the $c_9^{eff}$ term in Eqn.(\ref{eqn:eff}) gives zero contribution on contraction with the lepton bilinear, $c_7^{eff}$ gives zero by Eqn.(\ref{eqn:mesondecay2}) and the remaining $c_{10}$ gives a contribution given by
\bea \label{eqn:SMBsll}
i {M}_{SM} ({\overline B_s}(p_B) \to \mu^+(p_1) \mu^-(p_2)) &=&i {M}_{10} ({\overline B_s}(p_B) \to \mu^+(p_1) \mu^-(p_2)) \nonumber \\ 
&=& i \frac{G_F}{\sqrt{2}} V_{tb} V_{ts}^* c_{10} \langle \mu^+(p_1) \mu^-(p_2)| O_{10}| {\overline B_s}(p_B)\rangle,
\eea
where $O_{10}$ is given in Eq.~(\ref{eqn:effopr}). Accordingly the decay 
amplitude square is given by  
\bea \label{eqn:smamp}
{\overline{|{M}_{SM}|^2}} = {\overline{|{M}_{10}|^2}} .
\eea
For the explicit expression of ${\overline{|{M}_{10}|^2}}$ see Appendix A.   

\subsection{New Physics(NP) (moduli/radion contribution)}
The light stabilized radion ${\hat{\phi}(x)}$ can potentially be significant 
in the ${\overline B_s} \to \mu^+ \mu^-$ decay. Since gravitational interaction conserves flavour, 
radion can not cause flavour-changing neutral current (FCNC) transition 
$b\to s$ at the tree level and it can happen only at the loop level which we will see shortly. The radion interaction with the SM fields (which lives on the TeV brane) is governed by the 
$4$ dimensional general coordinate invariance. It couples to the trace of the
energy-momentum tensor $T^\mu_\mu (SM)$ of the SM fields and is given by 
\bea \label{eqn:radcoup}
{\mathcal{L}}_{\it int} = \frac{\hat{\phi}}{\vphi} T^\mu_\mu (SM).
\eea
Here $\vphi$ is the radion vev and $T^\mu_\mu (SM)$ is  
\bea \label{eqn:effint}
T^\mu_\mu (SM) = \sum_{\psi} \left[\frac{3 i}{2} \left({\overline{\psi}}
\gam_\mu D_\nu \psi - D_\nu {\overline{\psi}} \gam_\mu \psi \right)\eta^{\mu\nu}- 4 m_\psi {\overline{\psi}} \psi\right] - 2 M_W^2 W_\mu^+ W^{-\mu}
- m_Z^2 Z_\mu Z^\mu \nonumber \\
+ (2 m_h^2 h^2 - \partial_\mu h \partial^\mu h) + ...
\eea
where $D_\mu \psi = (\partial_\mu - i g \frac{\tau^a}{2} A_\mu^a - i g^\prime \frac{Y_W}{2} B_\mu) \psi$ ($\tau^a$'s being the Pauli matrices, $g,~g^\prime$ being $SU(2)_L$ and $U(1)_Y$ gauge coupling constants and $Y_W$, the weak hypercharge). Clearly
Eq.(\ref{eqn:effint}) manifests that radion intercation with matter conserves 
flavour at the tree level. The Feynman rules comprising radion interactions with the SM fields are listed in Appendix B.1 (see Figure 1).
For vertices involving radion-goldstone-goldstone, radion-goldstone-W boson, radion-fermion-antifermion-goldstone, radion-fermion-antifermion-W boson,  we read the Feynman rule for vertices 
\cite{ChengLi} without the radion and then just multiply the factor $1/\vphi$ to get the desired Feynman rule. For other vertices, the Feynman rules are directly read from Eq.~(\ref{eqn:radcoup}). 
Now a $b$ quark can decay to a $s$ quark together with a ultra-light radion($\phi$, dropping the hat from ${\hat \phi(x)}$ from now and onwards) 
followed by the $\phi \to l^+ l^-$ ($l=e, \mu$) decay. Since radion coupling to a pair of fermion (on-shell) is proportional to the fermion mass and $m_e = m_\mu \times 10^{-3}$, we do not
consider the $\phi \to e^+ e^-$ decay channel (and thus  ${\overline B_s} \to e^+ e^-$ decay mode) 
in our analysis, as it will not lead to useful bound on $\vphi$. 
At this point we should note that a  virtual radion of heavier
mass (say about few hundred GeV), might appears to be an interesting option. 
However, the possibility of FCNC transition (like $b \to s$ transition) with 
the radion, the real gravi-scalar, is ruled out. Note that an ultra-light radion of mass about $ 1 \sim 4$ GeV is not ruled out as far as the neutrino 
oscillaton inside supernova is concerned (see Ref.{\cite{SMahan}}). In Ref. 
\cite{SMahan}, the authors showed that the radion exchange (potential) 
between the neutrino and the supernova matter will not effect the neutrino oscillation inside the supernova, yielding a lower bound on $m_\phi$ about $1$ GeV. 
Another thing which requires attention is  the possibility of having the radion $\phi(x)$ mixing with the 
higgs $h(x)$ field. A term like
$$S = - \zeta \int d^4 x \sqrt{- g_{vis}}~ g_{vis}^{\mu \nu}~ H^\dagger H,$$ with
$H^\dagger = \left(0~~ (v + h(x))/\sqrt{2} \right)$ ( with $v=247$ GeV, the EW vev), can cause radion-higgs mixing \cite{Giudice}. However, here we confine ourselves in no-mixing scenario ( i.e. we set $\zeta$, the radion-higgs mixing parameter equal to zero). Analysis in a general radion-higgs mixing scenario 
is underway.
The Feynman diagrams contributing to the  $b \to s \phi$ FCNC transition are 
shown in Figure 2 (see Appendix B.2).

It is now straightforward to evaluate those diagrams and the effective operator parametrizing the radion contribution to $b\to s \phi$ process can be written as
\bea \label{eqn:effop}
O_{12} = V_{CKM}~\frac{G_F}{\sqrt{2}}~{\cal L}_\phi ~[m_b {\overline s}  P_R b + m_s  {\overline s} P_L b],
\eea
where  
$V_{CKM} = V^*_{is} V_{ib} (i=u,c,t)$, $G_F (= \frac{g^2 \sqrt{2}}{8 M^2_W})$, the Fermic constant. ${\cal L}_\phi$, the effective loop 
integral factor takes the form  
\bea \label{eqn:effloopfac}
{\cal L}_\phi = \frac{m_i^2 }{2 \pi^2 \vphi} \int_0^1 d x \left[4~{Log\left(\frac{\Lambda^2}{A^2}\right) - 4} + x \frac{m_i^2}{A^2} + \frac{(1 -x) M_W^2}{A^2}\right],
\eea
in the t'Hooft-Feynman gauge $\xi = 1$ and $A^2 = x m_i^2 + (1- x) M_W^2$. 
In Eq.~(\ref{eqn:effloopfac})
$i$ runs over $u$, $c$ and $t$ quarks, but in our analysis we set $i = t$ (since $m_t \gg m_u, m_d$).  
In evaluating ${\cal L}_\phi$ we assume that there is no external momentum flow (actually the external momenta is much smaller than the internal masses 
$M_W$ and $m_t$ and hence setting the external momentum to zero is a 
good approximation). 
Note that, the  Figures 2(b,e,i,j) contribute in $b \to s \phi$ vertex (Eq. (\ref{eqn:effloopfac})), while the rest are not. We use the cut-off
regularizing technique in regularizing the loop integral with the UV cut-off 
being 
$\Lambda = 4 \pi \vphi$ (follows from naive-dimensional analysis). 

\begin{figure}[htb]
\begin{center}
\vspace*{1.0in}
      \relax\noindent\hskip -4.4in\relax{\includegraphics{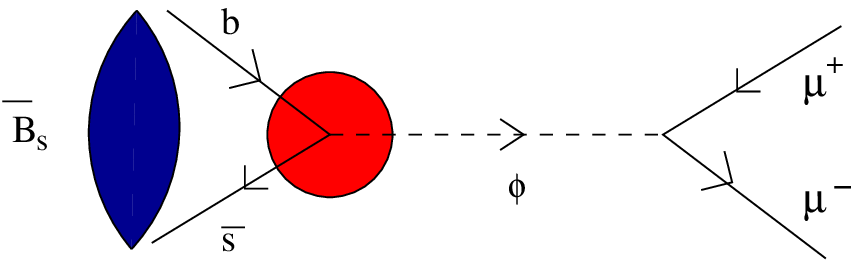}}
\end{center}
\end{figure}
\vspace*{0.2in}
\noindent {Figure 3}.
{ \it Radion contribution to ${\overline B_s}(p_B) \to \mu^+(p_1) \mu^-(p_2) decay$. Momentum conservation (for the semileptonic process $b(p_b) {\overline s}(p_s) \to \mu^+ (p_1) \mu^- (p_2)$) reads $p_B = p_b + p_s = p_1 + p_2$. 
The red blob corresponds to the effective $b-{\overline s}-\phi$ vertex.}      

 In finding the NP (radion) contribution to ${\overline B_s} \to \mu^+ \mu^-$ 
 decay, we neglect the $m_s$ dependent term in Eq.(\ref{eqn:effop}) (since $m_s \ll m_b$).
It is now straightforward to include the NP contribution to the 
${\overline B_s} \to \mu^+ \mu^-$ decay amplitude. The relevant new physics 
Feynman digram is shown in Figure 3 and  
the squared-amplitude ~${\overline{|M_{SM} + M_{NP}|^2}}({\overline B_s} \to \mu^+ \mu^-)$ is given by
\bea \label{eqn:NPBsll}
{\overline{|M_{SM} + M_{NP}|^2}} = {\overline{|M_{10}|^2}} + {\overline{|M_{NP}|^2}} + {\overline{2 Re(M_{10}^\dagger  M_{NP}})} .
\eea
The direct and interference terms of the amplitude (Eq.(\ref{eqn:NPBsll})) is presented in the Appendix A and B.3. Note that, in Figure 3 the radion $\phi$, produced in the annihilation of $b$ and ${\overline s}$ quarks, decays into 
$\mu^+ \mu^-$ pairs. We are working in the Breit-Wigner approximation and in this approximation the radion propagator takes the form  
\bea
\Delta_\phi = \frac{1}{\left[(q^2- m_\phi^2) + i m_\phi \Gamma_\phi \right]}.
\eea
Here $q^2= (p_b + p_s)^2 = (p_1 + p_2)^2$ and $\Gamma_\phi$ is the total width of the radion arising from it's decay $\phi \to f{\overline f} (f=u,d,c,s,t,b,e,\mu,\tau,\nu_e,\nu_\mu,\nu_\tau),~WW,ZZ,hh$ (for the radion decay width and it's branching ratio to several channels see Ref. \cite{DasSreerup}).

Considering the NP contribution, the total decay width 
$\Gamma({\overline B_s} \to \mu^+ \mu^-)$ can be written as
\bea \label{eqn:SMBslplm}
  \Gamma ( {\overline B_s} \rightarrow \mu^+ \mu^- ) =  \frac{p_c}{8 \pi m_{{\overline B_s}}^2} {\overline{|{M}_{TOT}|^2}},
  \eea
   where ${\overline{|M_{TOT}|^2}} = {\overline{|M_{SM} + M_{NP}|^2}}$ (as given above (Eq.~\ref{eqn:NPBsll})).  $p_c$ is the center of mass (c.o.m) momenta of the two charged muons in the ${\overline B_s}$ rest frame and is given by
   \bea
   p_c = \frac{\sqrt{\left(m_{{\overline B_s}}^2 - (m_{\mu^+} - m_{\mu^-})^2\right)
   ~\left(m_{{\overline B_s}}^2 - (m_{\mu^+} + m_{\mu^-})^2\right)}}{2 m_{{\overline B_s}}}.
   \eea
Finally the branching ratio $BR({\overline B_s} \to \mu^+ \mu^-)$ is given by 
   \bea \label{eqn:BRtot}
   BR({\overline B_s} \to \mu^+ \mu^-) = \tau_{{\overline B_s}} \Gamma ({\overline B_s} \to \mu^+ \mu^-),
   \eea
   where $\tau_{{\overline B_s}} (= 1.461 \times 10^{-12}~ sec $) is the life-time of the ${\overline B_s}$ meson \cite{PDG}.
   
\section{Input parameters} \label{sec:section3}
The decay amplitudes depend on the 
CKM matrix elements, wilson coefficients, quark and lepton masses and the non-perturbative input like decay constants.
\subsection{CKM matrix elements, quark masses, wilson coefficients and decay constant}
We adopt the Wolfenstein parametrization with parameters $A,
\lam, \rho$ and $\eta$ of the CKM matrix as below
\bea
V_{CKM} =
\left( \begin{array}{ccc}
V_{ud} & V_{us}  & V_{ub} \\
V_{cd} & V_{cs}  & V_{cb} \\
V_{td} & V_{ts}  & V_{tb}\end{array} \right)
=\left(\begin{array}{ccc}
1 - \frac{1}{2}\lam^2 & \lam & A \lam^3 (\r - i \e) \\
-\lam & 1 - \frac{1}{2}\lam^2  & A \lam^2 \\
A \lam^3 (1 - \r- i \e) & - A \lam^2 & 1\end{array} \right).
\eea
 We set $A$ and $\lam = \sin\theta_c$ ($\theta_c$, the Cabibbo mixing angle) 
at $0.801$ and $0.2265$ respectively, in our analysis. Other relevant parameters
are $ \r = \sqrt{{\overline \r}^2 + {\overline \eta}^2} 
~cos{\gamma}$ and $ \e = \sqrt{{\overline \r}^2 + {\overline \eta}^2}~ 
sin{\gamma}$, where $\sqrt{{\overline \r}^2 + {\overline \eta}^2}
= 0.4048$ and $\gamma \simeq 62^{o}$ \cite{CKMF}.
The quark masses(in GeV unit) are being set at their current values i.e. $m_u=0.2,~m_d=0.2,~m_s=0.2,~m_c=1.4,~m_b=4.8,~m_t=175$ GeV \cite{Ali} and $m_\mu = 0.105$ GeV. For the ${\overline B_s}$ meson 
we use $m_{{\overline B_s}} = 5.369$ GeV \cite{PDG} and the decay constant $f_{{\overline B_s}} = 0.20$~GeV \cite{CY}. For our numerical calculation we use $c_{10}=-4.5461$ \cite{Ali}.

\section{Numerical Analysis: ~Results and Discussions}
\label{sec:section4}

The branching ratio $BR({\overline B_s}\to \mu^+ \mu^-)_{SM}$ in the Standard
Model is found to be two order smaller \cite{Buras} than the
present experimental upper bound  $1.0 \times 10^{-7}$ \cite{HFAG}.
Assuming that the future data (not the upper bound) will still differ from the SM result, it is thus worthwhile to see whether radion can explain the discrepency. Our analysis is organized as follows:

We first obtain the SM result. Using several numerical inputs (discussed in section 3), we find
 \bea
  BR({\overline B_s}\to  \mu^+ \mu^-)_{SM} = 3.17 \times 10^{-9},
 \eea

\vspace*{0.15in}
\noindent which is two order smaller than the present experimental upper
bound. Next we will see whether the radion can explain the above deficit or not.Taking into account the radion contribution along with the SM contribution,
we find the total branching ratio which is given by Eq.~(\ref{eqn:BRtot}) and
is solely a function of the new physics parameters: (1) radion mass $m_\phi$ and (2) radion vev $\vphi$. In our analysis, we treat the radion as an
ultra-light object whose mass $m_\phi$ varies in between $2 m_\mu$ GeV to
$m_{B_s}$ GeV and in this mass range it's decay width is found to be
quite small. We set it at $\Gamma_\phi=0.001$ GeV (see \cite{DasSreerup} for an elaborate discussion on radion decay width).\\

Now let us define a quantity $R = \frac{BR({\overline B_s} \to \mu^+ \mu^-)}{BR({\overline B_s} \to \mu^+ \mu^-)_{SM}}$ which is a function of
$m_\phi$ and $\vphi$, the two NP parameters(as mentioned above).
$R$, which is purely to be determined from the experiment, can be potentially
quite significant in radion discovery.

Since at present no data for $BR({\overline B_s} \to  \mu^+ \mu^-)$ is
available (only upper bound exists), we take the conservative viewpoint and we
use $R$ to impose constraints on $m_\phi$ and $\vphi$. What we do is as follows:
\begin{itemize}
\item We set $R =10,~50$ and $100$ and use those to obtain contour plots in
the $m_\phi-\vphi$ plane, which are shown in Figure 4a. The topmost, middle and lowermost curves respectively stands for $R=10, 50$ and $100$.
In Figure 4b we have plotted $R$ as  a function of $m_\phi$ and
$\vphi$.

\item From Figure 4a we can see that the lower bound on $\vphi$ for a
given $m_\phi$ decreases with the increase in $R$.
For $R =10(50, 100)$ and $m_\phi= 2$ GeV, we find $\vphi \ge 918(564, 459)$ GeV.
\end{itemize}

\newpage
\vspace*{-1.0in}
\begin{figure}
\subfigure[]{
\label{PictureOneLabel}
\hspace*{-0.55in}
\begin{minipage}[b]{0.5\textwidth}
\centering
\includegraphics[scale=1.1]{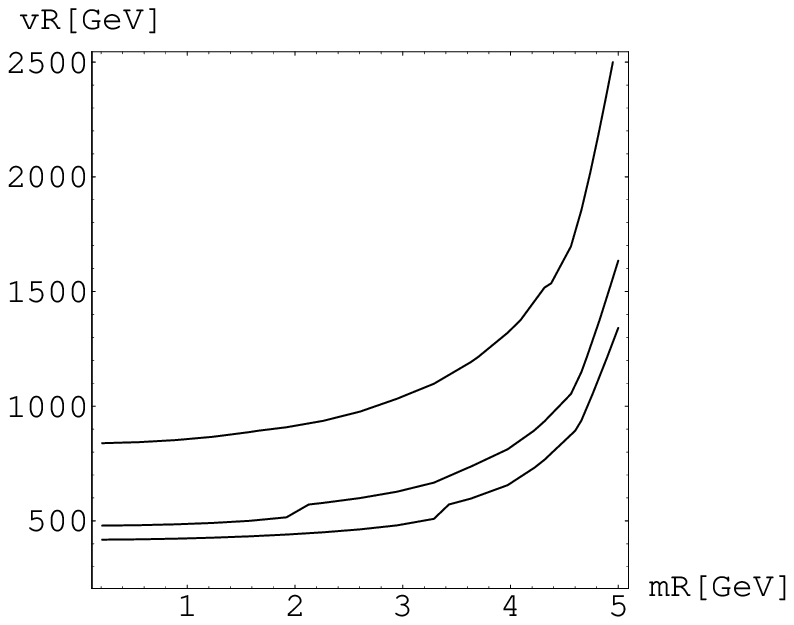}
\end{minipage}}
\subfigure[]{
\label{PictureTwoLabel}
\hspace*{0.3in}
\begin{minipage}[b]{0.5\textwidth}
\centering
\includegraphics[scale=1.1]{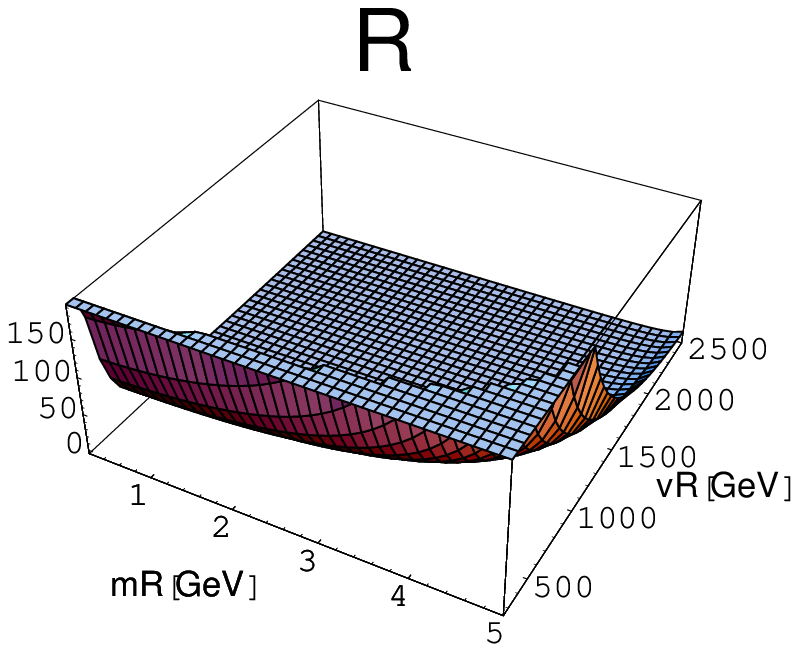}
\end{minipage}}
\end{figure}
\vspace*{0.5in}
\noindent {\bf Figure 4(a,b)}.
{{\it Figure 4a corresponds to the contour plots in the $mR (m_\phi)$ -$vR 
(\vphi)$ plane. Here $mR$ and $vR$ are in GeV unit. The contour line starting from the top to bottom corresponds to $R = 10,~50$ and $100$, respectively. In Figure 4b, we have shown the variation of $R$ with $mR$ and $vR$.}}
\begin{itemize}
\item For a given $R$ curve, we see that the lower bound on $\vphi$ increases
with $m_\phi$. For example, consider the $R=10$ curve of Figure 4a. As
 $m_\phi$ varies from $0.21$ GeV to $5$ GeV, $\vphi$ (the lower bound)
changes from $844$ GeV to $2620$ GeV. Similarly for the $R=100$ curve and for
the above $m_\phi$ range, the lower bound on $\vphi$ varies from $421$
GeV to $1342$ GeV.

\item In Figure 5 we have plotted  $BR({\overline B_s} \to \mu^+ \mu^-)$
as a function of $\vphi$ for $m_\phi = 1$ GeV (lower curve) and $4$ GeV
(upper curve), respectively. The horizontal line, the present experimental upper bound($= 1.0 \times 10^{-7}$), suggest that the region allowed for $\vphi$ lies below the horizontal curve. This immediately allows one to obtain lower bound on $\vphi$ and we find
for $m_\phi = 1(4)$ GeV, the lower bound is about $605(945)$ GeV.

\end{itemize} 
\newpage
\begin{figure}[htb]
\begin{center}
\vspace*{1.5in}
      \relax\noindent\hskip -5.4in\relax{\includegraphics{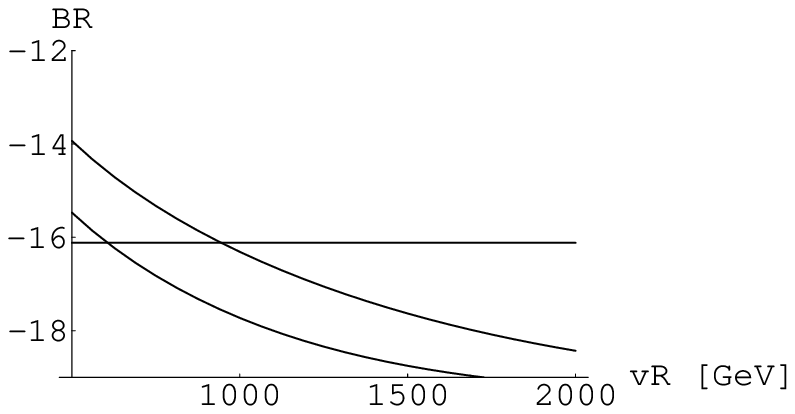}}
\end{center}
\end{figure}
\vspace*{-0.25in}
\noindent {\bf Figure 5}.
{{\it In Figure 5, the $BR({\overline B_s} \to \mu^+ \mu^-)$ as a function of
$vR(\vphi)$ (GeV) is shown. In y-axis we have actually plotted $Log[BR({\overline B_s} \to \mu^+ \mu^-)]$. For example, $BR = -18(-14)$ corresponds to 
$1.523 \times 10^{-8} (8.315 \times 10^{-7})$. The lower(upper) curve 
corresponds to $m_\phi= 1(4)$ GeV, respectively. The horizontal curve 
(corresponding to $BR({\overline B_s} \to \mu^+ \mu^-) = 1.0 \times 10^{-7}$) represents the present experimental upper bound. The region below the horizontal curve is allowed for vR($\vphi$). }}

\vspace*{0.15in}
\noindent {\large {\bf{Invariant mass distribution}}}\\

\hspace*{0.1in} The invariant mass distribution is an useful probe to see the NP signal and we now obtain such a distribution to see the radion signal.
 In Figure 6, we have plotted $d \Gamma/d M_{\mu \mu}$ as a function of the 
 $M_{\mu \mu}$ (the di-muon invariant mass), respectively for $\vphi = 247,~500$ GeV and $1$ TeV (from top to bottom). The lowermost line corresponds to the SM background. Clearly for all  
 $\vphi$ ranging from $247$ GeV to $1$ TeV, the signal is way above the SM 
 background and the height of the resonance peak increases. To see these consider the curves corresponding to $m_\phi=1$ GeV and $\vphi = 247,500$ and $1000$ GeV. One finds $d \Gamma/d M_{\mu \mu}$ as $\sim 5.5 \times 10^{-13}$, $8.55 \times 10^{-13}$ and $1.03 \times 10^{-12}$, while for the SM one finds $d \Gamma/d M_{\mu \mu}=4.62 \times 10^{-23}$. 
 The detectibility ofcourse
  depends on the clarity of the peak structure.
 Since the decay width goes as $1/\vphi^2$, with the increase of $\vphi$, the resonance width decreases, one can have the better chance to see the resonance due to radion. We also see that with the increase of $\vphi$, resonance of higher radion mass gets lost and at $\vphi = 1$ TeV, resonance corresponding to $m_\phi = 1,~2$ GeV persists. We note that for $\vphi = 2$ TeV, the resonance corresponding to $m_\phi = 1$ GeV only survives (which is not shown in the figure).
\newpage
\vspace*{3.78in}
\begin{figure}[htb]
\begin{center}
\vspace*{4.5in}
      \relax\noindent\hskip -6.4in\relax{\includegraphics{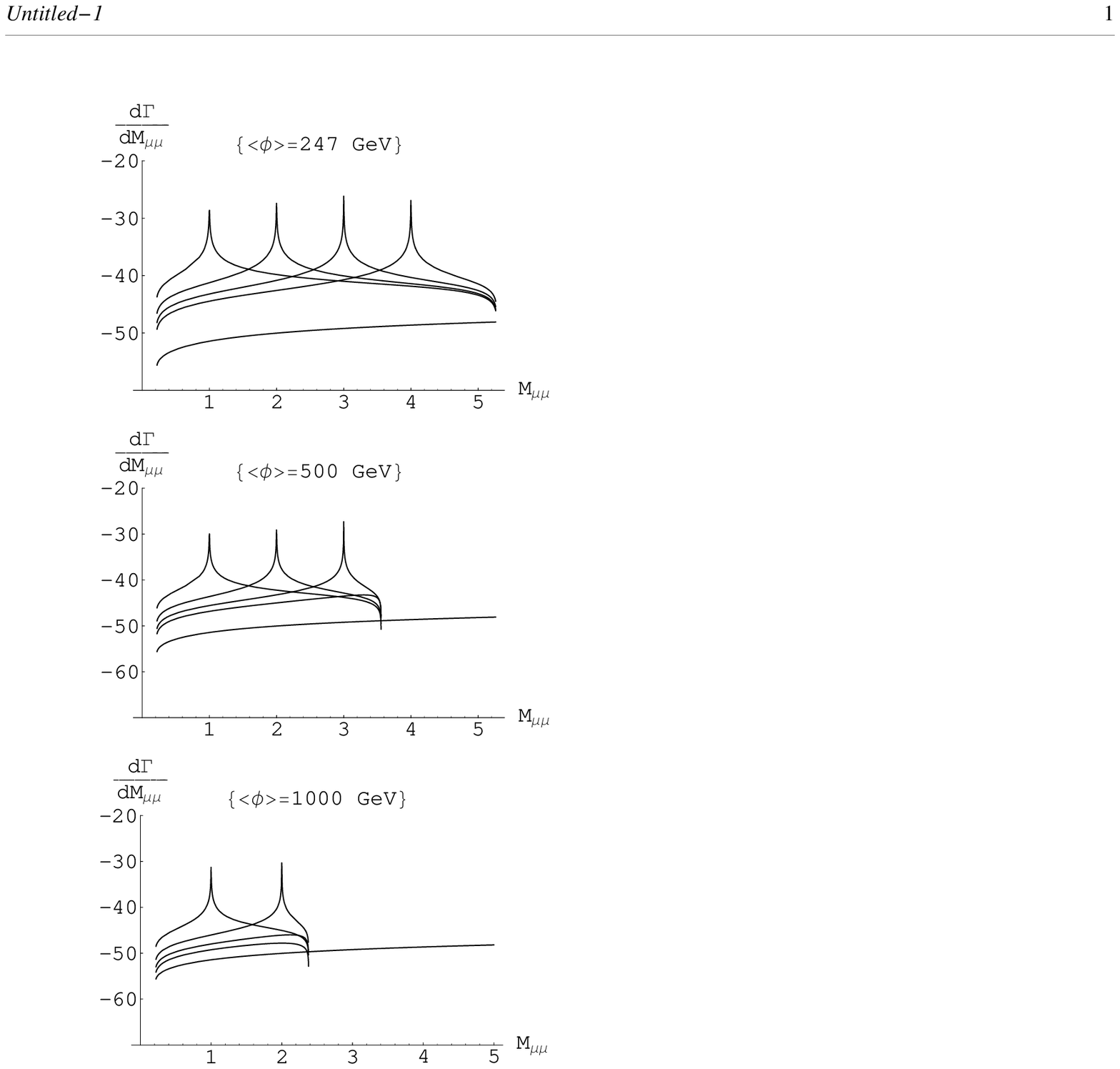}}
\end{center}
\end{figure}
\vspace*{-4.0in}
\noindent {Figure 6}.
{ \it Invariant mass distribution $d \Gamma/d M_{\mu\mu}$ for the ultra-light
radion signal with $m_\phi = 1,2, 3$ and $4$ GeV (from left to right for a given curve), where the lowermost curve corresponds to the SM background. In y-axis we have actually plotted $Log[d\Gamma({\overline B_s} \to \mu^+ \mu^-)/d M_{\mu\mu}]$. The topmost,middle and lowermost curves
corresponds to $\vphi = 247,~500$ GeV and $1$ TeV, respectively.}

\newpage
\noindent  {\large {\bf  Comparison with other studies}}\\

\hspace*{0.1in} We next compare our bound on $\vphi$ with that obtained from 
the muon anomaly data.  The present BNL 
muon anomaly data tells $a_\mu^{(expt)} - a_\mu^{(SM)}= (28 \pm 10.5)\times 10^{-10} $. Fitting it 
with $\delta a_\mu^{\phi}$ (radion contribution to muon anomaly) the author 
in \cite{DasP} found the following bound for a relatively light radion:
for $m_\phi = 2$ GeV, the lower bound on $\vphi$ is $\sim 500$ GeV 
corresponding to $\delta a_\mu^{\phi} = 3.85 \times 10^{-9}$ and
$\sim 760$ GeV for $\delta a_\mu^{\phi} = 1.75 \times 10^{-9}$.
The upper bound on $BR({\overline B_s} \to \mu^+ \mu^-)$ suggests a lower bound  about $918$ GeV (corresponding to $R = 10$) $\sim 459$ GeV (corresponding to $R = 100$) for $m_\phi = 2$ GeV and is in agreement with that obtained from the 
muon anomaly data as mentioned above.
	
\section{Summary and Conclusion} \label{sec:section5}
We analyse the ${\overline B_s} \to \mu^+ \mu^-$ decay in the Randall-Sundrum model in the presence of an ultra-light stabilized radion. 
Using the present experimental bound of the branching ratio 
$BR({\overline B_s} \to \mu^+ \mu^-)$ 
(two order larger than the standard model result), 
we obtain contour plots in the $m_\phi - \vphi$ plane.
For $BR({\overline B_s} \to \mu^+ \mu^-) = 1.0 \times 10^{-7}$ and 
radion mass $m_\phi = 0.21(5)$ GeV, we obtain the lower bound on $\vphi$ 
about $844(2620)$ GeV corresponding to $R=10$ and about $421(1342)$ GeV corresponding to $R=100$. We have shown the invariant mass ($M_{\mu\mu}$) 
distribution plot which clearly shows the existence of the radion signal. The SM background  on which the radion signal lies is highly suppressed. 
Finally, we compare our result
with the bound obtained from the BNL muon anomaly data and found some agreement between the two.  
\section{Acknowledgments}
We would like to thank Prof.R.~Sinha of IMSc, Chennai for correcting our 
mistakes and  his several useful comments related to this work. P.K.D would like to thank Prof.Uma Mahanta(late) who first introduced the author to the beauty of the Brane World Physics.  

\newpage
\appendix
\section{SM amplitude of the process ${\overline B_s} \rightarrow \mu^+ \mu^-$}

In this appendix, we calculate the square of the SM amplitude of 
${\overline B_s} \rightarrow \mu^+ \mu^-$ (at the quark level it yields $b {\overline s} \to \mu^+ \mu^-$ process) of Eq.(\ref{eqn:smamp}). We define 
$p_b$, $p_s$, $p_1$ and $p_2$ to be the momenta of the b-quark,
s-quark, $\mu^+$ and $\mu^-$, respectively and $m_\mu$, the muon mass and  $q = p_b +p_s = p_1 + p_2$.
The individual amplitude-square elements are listed below:
\bea
\overline {|{\cal M}_{10}|^{2}}&=& C' |C_{10}|^{2}[ 4{m_{\mu}}^2 {m_b}^2 - 
4 (p_1.p_2){m_b}^2 + 4{m_{\mu}}^2 {m_s}^2 -4 (p_1.p_2){m_s}^2 \nonumber\\
& & +8(p_1.p_b)(p_2.p_b) +8(p_1.p_s)(p_2.p_b)+8(p_1.p_b)(p_2.p_s)+
8(p_1.p_s)(p_2.p_s)\nonumber\\
& & +8 {m_{\mu}}^2 (p_b.p_s) - 8(p_1.p_2)(p_b.p_s)].
\eea
where the prefactor $C'$ is defined as follows
\bea
C' = (\frac{\alpha G_{F}f_{{\overline B_s}}}{2 \pi\sqrt{2}})^{2} |V_{tb}|^{2}
|V_{ts}|^{2}.
\eea


\newpage
\section{ NP contribution}
\subsection{Interaction vertex of Radion and SM particle}
\begin{figure}[htbp]
\centerline{\hspace{-3.3mm}
{\epsfxsize=16cm\epsfbox{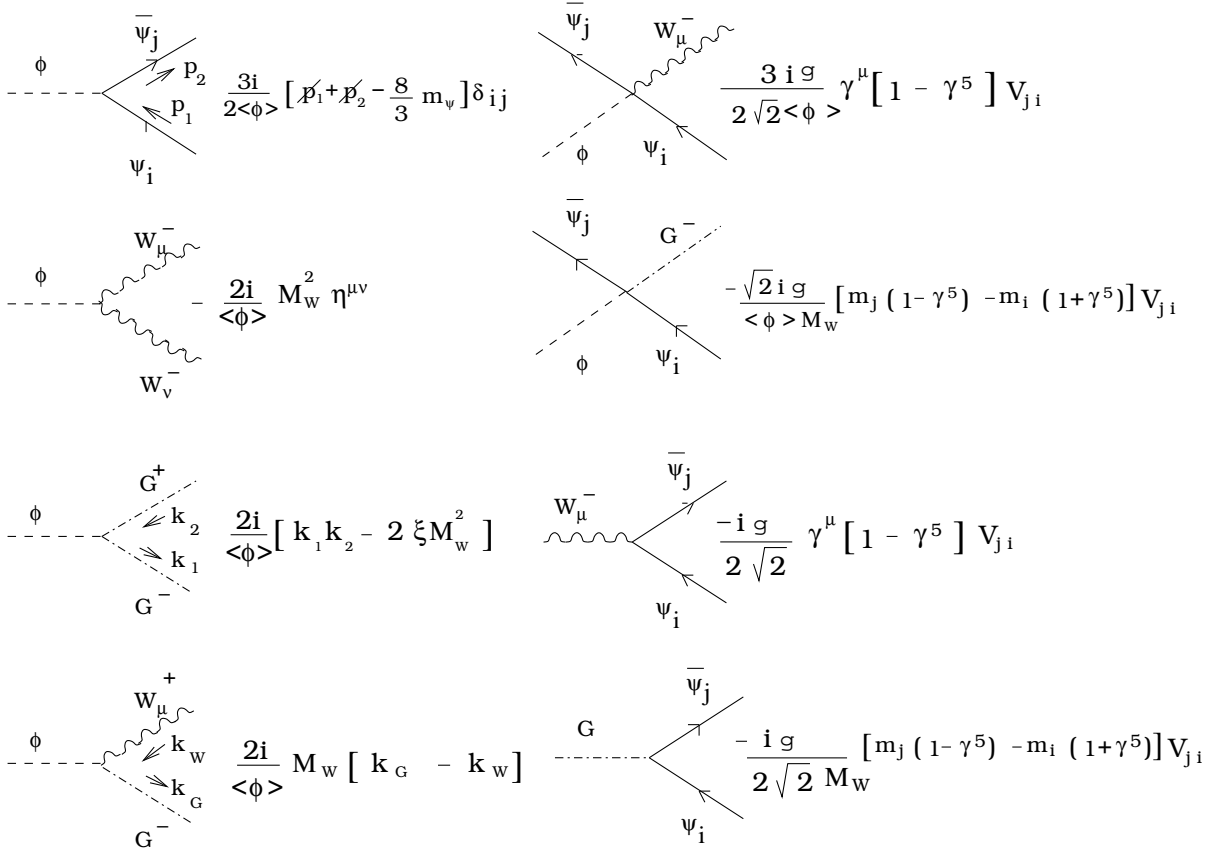}}}
\hspace{3.3cm}
\caption{The relevant diagrams of the interaction vertex of Radion 
and SM particles in the $R_\xi$ gauge.}
\protect\label{fig1}
\end{figure}
\newpage
\subsection{Feynman diagrams of the Radion penguin with SM particles.}
\begin{figure}[htbp]
\centerline{\hspace{-3.3mm}
{\epsfxsize=16cm\epsfbox{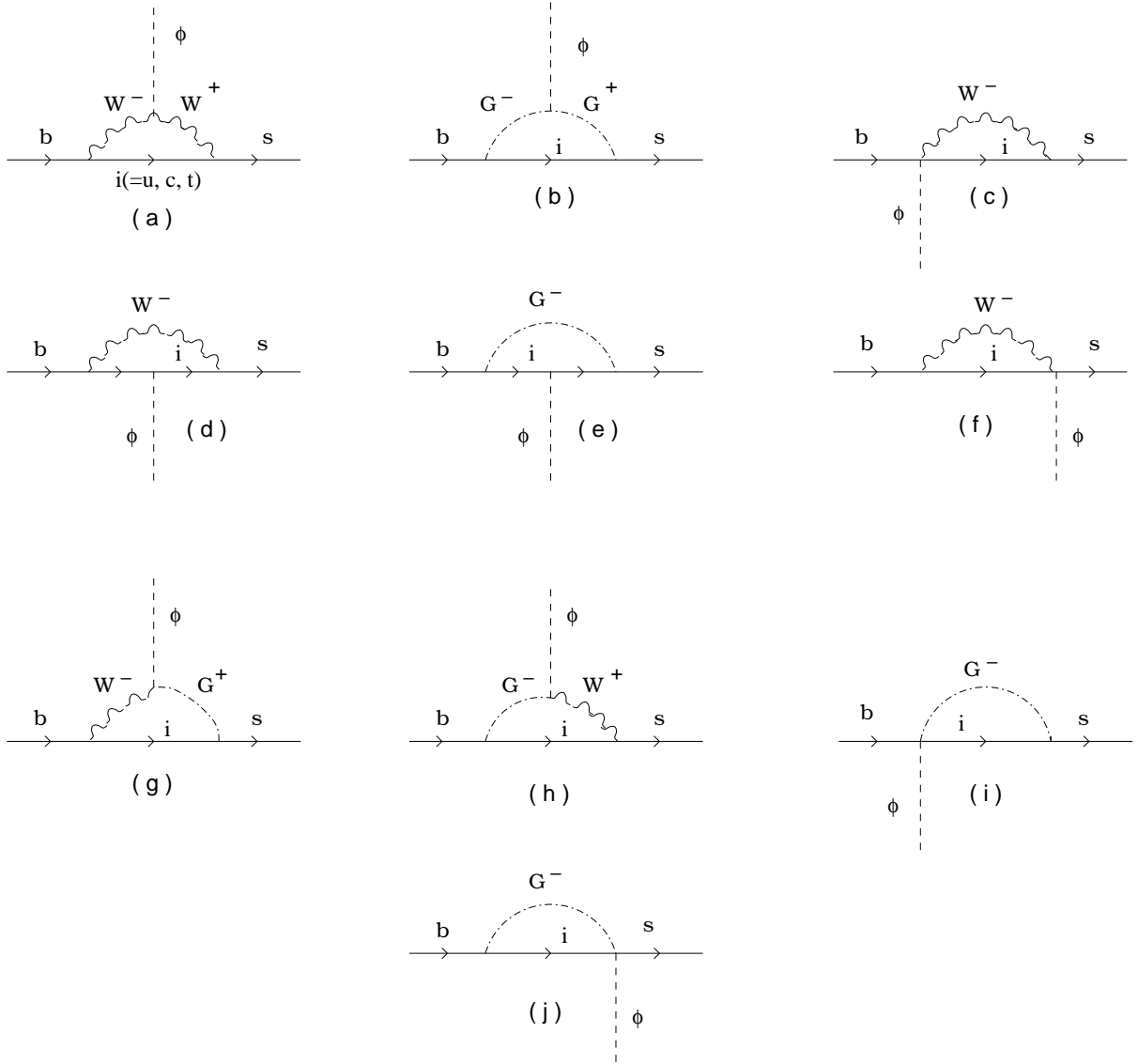}}}
\hspace{3.3cm}
\caption{The Feynman diagrams of the Radion penguin with SM particles
in the general $R_\xi$ gauge .}
\protect\label{fig2}
\end{figure}
\newpage
\subsection{NP amplitude of the process ${\overline B_s} \rightarrow  \mu^+ \mu^-$}
In this appendix, we calculate the square of the NP amplitude of 
${\overline B_s} \rightarrow  \mu^+ \mu^-$ of Eq.(\ref{eqn:NPBsll}).
The individual amplitude-squared elements are given by  
\bea
\overline {|{\cal M}_{NP}|^{2}}&=&  C_L (p_1.p_2 -{m_\mu}^2),\\
\overline {2Re({\cal M}_{10}^{\dagger}{\cal M}_{NP})} &=& 0,
\eea
where  
\bea
 C_L&=& |V_{tb}|^2|V_{ts}|^2 \left(\frac{G_F m_b m_\mu {\cal L}_\phi}
 {\sqrt{2} \vphi }\right)^2 \left( \frac{ {f_{{\overline B_s}}}{m_{{\overline B_s}}}^2 }{m_b + m_s}
 \right)^2 
\times \frac{1}{ (q^2 - m_\phi^2)^2 + {\Gamma}_\phi^2 m_\phi^2}, \nonumber \\ 
{\cal L}_\phi &=& \frac{ {m_t}^2}{2 {\pi}^2 \vphi}{\int_{0}}^{1}F(x)dx.
\nonumber 
\eea
In above
\bea
F(x) &=& 4(Log(\Lambda^2/A^2)-1)+\frac{x{m_t}^2}{A^2} + \frac{(1-x)
 {M_W}^2}{A^2}, \nonumber \\
\Lambda &=& 4 \pi \vphi,~~ A^2= x {m_t}^2 +(1-x){M_W}^2.\nonumber  
\eea
We worked in the `t-Hooft Feynman gauge ($\xi = 1$).


\end{document}